\documentclass[12pt]{iopart}

\usepackage[dvips]{graphicx}

\newcommand{\Gcenter}[2]{
	\dimen0=\ht\strutbox%
	\advance\dimen0\dp\strutbox%
	\multiply\dimen0 by#1%
	\divide\dimen0 by2%
	\advance\dimen0 by-.5\normalbaselineskip
	\raisebox{-\dimen0}[0pt][0pt]{#2}
}

\begin{document}

\title[Identification problems of $\mu$ and $e$ events in the Super K.]{Identification problems of muon and electron events in the Super-Kamiokande detector} 

\author{K Mitsui\dag, T Kitamura\ddag, T Wada\S, and K Okei\S}
\address{\dag Yamanashi Gakuin University, Kofu 400-8575, Japan}
\address{\ddag c/o RIST, Kinki University, Higashi-Osaka 577-8502, Japan}
\address{\S Department of Physics, Okayama University, Okayama 700-8530, Japan}  
\ead{twada@science.okayama-u.ac.jp}

\begin{abstract}
 In the measurement of atmospheric $\nu_e$ and $\nu_{\mu}$ fluxes, 
the calculations of the Super Kamiokande group for the distinction 
between muon-like and electron-like events observed 
in the water \v{C}erenkov detector have initially assumed 
a mis-identification probability of less than 1 \%  and later 2 \% 
for the sub-GeV range. 
In the multi-GeV range, they compared only the observed 
behaviors of ring patterns of muon and electron events, 
and claimed a 3 \% mis-identification.
However, the expressions and the calculation method 
do not include the fluctuation properties due to the stochastic nature 
of the processes which determine 
the expected number of photoelectrons (p.e.) produced by muons and electrons.
Our full Monte Carlo (MC) simulations 
including the fluctuations of photoelectron production show that 
the total mis-identification rate for electrons and muons 
should be larger than or equal to 20 \% for sub-GeV region.  
Even in the multi-GeV region we expect a mis-identification rate 
of several \%  based on our MC simulations taking into account 
the ring patterns. 
The mis-identified events are mostly of muonic origin. 
    
Comparing the intensities per kiloton-year between the 33 kiloton-year 
(the first half of the total exposure of 79.3 kty) 
and 46.3 kton-year (the latter half) exposures, moreover, 
the respective values for $e$-like event numbers, 
$\mu$-like event numbers and the ratio for 
($e$-like number)/($\mu$-like number) are not consistent 
within the given statistical uncertainties 
in the sub-GeV and multi-GeV regions for the two subsamples. 
It should be also remarked that results for 
the whole zenith angle distributions of e-like and of $\mu$-like events 
in the sub-GeV ranges and those of $e$-like and 
$\mu$-like(FC) events in the multi-GeV ranges 
are inconsistent for the 33 kton-year and 46.3 kton-year exposures. 
We also studied the significance probability of 
the results after subdividing the whole zenith angle distribution 
into upward ($-1 \le \cos \theta \le 0$) 
and downward ($0 \le \cos \theta \le 1$) going events. 
Their significance probabilities also support our claim 
that the method used by the Super K. group is inadequate for 
the identification of muons and electrons.
\end{abstract}

\pacs{95.55.Vj,96.40.Tv}

\submitto{\jpg}

\maketitle

 \section{Introduction}
The most recent Super Kamiokande (Super K.) results for 
the study of atmospheric neutrinos and the neutrino anomaly 
before  the large accident of the detector have been presented 
for the 79.3 kiloton-year (kty) exposure by Kajita and 
 Totsuka~\cite{TKYT} and at the 27th International Cosmic Ray Conference, 
Hamburg, August 2001~\cite{JKa}. 
Using their reports, we have compared 
the respective intensities/kty as shown in Table 1 
for the sub-GeV results (visible energy, $E_{vis}$ \textless 1330 MeV) 
and for the multi-GeV results ($E_{vis}$ \textgreater 1330 MeV). 
These results are subdivided into exposures of 25.5 kty(analysis A) 
and 25.8 kty(analysis B)~\cite{YF,YFuk}, 
also 33.0 kty~\cite{YFuku,YSu,KK} 
and 46.3 ($= 79.3 - 33.0$) kty~\cite{TKYT,JKa}, 
as given in the published papers and reports.

Table \ref{table:table1} shows a comparison of various event samples 
for $e$-like event numbers, $\mu$-like event numbers, ($e$-like+$\mu$-like) 
numbers and the ratio for ($e$-like number)/($\mu$-like number) 
in the sub-GeV and multi-GeV regions~\cite{KMi}. 
The actual event numbers are shown in the second and the fourth column. 
Disagreements between the related two exposures are indicated 
by respective identifier marks. 
Earlier data for the sub-GeV region collected 
during the period from May 1996 to October 1997 
have been analysed independently by group A for the exposure of 25.5 kty 
and group B for the exposure of 25.8 kty, respectively~\cite{YF}. 
The comparisons are shown in the upper part in table \ref{table:table1}. 
Both the event rates per kty for $\mu$-like number, 
($e$-like number + $\mu$-like number) and the ratio for 
($e$-like number)/($\mu$-like number) appear to be in conflict above 
the respective  statistical uncertainties. 
Therefore, our main purpose for the present considerations is 
to compare those event rates for both exposures. 
The systematic uncertainties depend on the different kty values, 
but they are smaller than the statistical ones. 
 
In both A and B groups, 
the event rates for $\mu$-like event numbers marked by $\triangleleft$ 
and for ($e$-like + $\mu$-like) event numbers marked by $\triangleright$ 
in the sub-GeV range do not agree. 
Also, the ratios for ($e$-like)/($\mu$-like) events marked by $\triangle$ are 
inconsistent in spite of the fact that the same data sample is analysed. 
However, the Super K. group finally has considered only 
the result of analysis A disregarding analysis B 
which has given larger number of $\mu$-like events than analysis A. 

\begin{table}[htbp]
\begin{center}
\caption{Comparisons of various event samples for sub-GeV and multi-GeV in the Super Kamiokande detector.}
\label{table:table1}
\small
\begin{tabular}{|l|r|cr@{.}l@{${}\pm{}$}r@{.}lc|r|cr@{.}l@{${}\pm{}$}r@{.}lc|}
\hline \hline 
{\bf Sub-GeV} & 
\multicolumn{7}{|c|}{{\bf 25.5 kty (Analysis A)}} 
& \multicolumn{7}{|c|}{{\bf 25.8 kty (Analysis B)}} \\ \hline \hline
$e$-like number & 
983 & &38&6&1&2&/kty& 
967 & &37&5&1&2&/kty\\
$\mu$-like number & 
900 &$\triangleleft$&35&3&1&2&/kty&
1041 &$\triangleleft$&40&4&1&3&/kty\\
$e$-like+$\mu$-like&
1883 &$\triangleright$&73&84&1&70&/kty&
2008 &$\triangleright$&78&75&1&76&/kty\\
($e$-like)/($\mu$-like)& 
&$\bigtriangleup$&1&09&0&05& &
&$\bigtriangleup$&0&93&0&04& \\ \hline \hline
{\bf Sub-GeV} & 
\multicolumn{7}{|c|}{{\bf 33.0 kty ($\nu$-98)}} 
& \multicolumn{7}{|c|}{{\bf 46.3 kty} (=79.3-33.0 kty)} \\ \hline \hline
$e$-like number & 
1231 &$\bigcirc$&37&30&1&06&/kty& 
1633 &$\bigcirc$&35&27&0&87&/kty\\
$\mu$-like number & 
1158 &&35&09&1&03&/kty&
1328 &&35&21&0&87&/kty\\
$e$-like+$\mu$-like&
2389 &&72&39&1&48&/kty&
3263 &&70&48&1&23&/kty\\
($e$-like)/($\mu$-like)& 
&&1&06&0&04& &
&&1&00&0&04& \\ \hline \hline
{\bf Multi-GeV} & 
\multicolumn{7}{|c|}{{\bf 33.0 kty ($\nu$-98)}} 
& \multicolumn{7}{|c|}{{\bf 46.3 kty} (=79.3-33.0 kty)} \\ \hline \hline
$e$-like number & 
290 &$\otimes$&8&8&0&5&/kty& 
334 &$\otimes$&7&2&0&4&/kty\\
$\mu$-like(FC) number & 
230 &&7&0&0&5&/kty&
328 &&7&1&0&4&/kty\\
$e$-like+$\mu$-like(FC)&
520 &$\ominus$&15&8&0&7&/kty&
662 &$\ominus$&14&3&0&6&/kty\\
($e$-like)/($\mu$-like)& 
&$\odot$&1&26&0&11& &
&$\odot$&1&02&0&08& \\ \hline 
$\mu$-like(PC) number&
301&&9&1&0&5&/kty&
453&&9&8&0&5&/kty \\ \hline \hline
\end{tabular}
\normalsize
\end{center}
\end{table}    
 
Various observed intensity values of the 33.0 kty exposure 
(including the data in 25.5 kty run~\cite{YF}) are compared to 
those of the 46.3 kty~\cite{TKYT} in the central part of Table 1. 
The former is the first half in the total 79.3 kty exposure 
and the other is the latter half of the 79.3 kty exposure. 
The $e$-like intensities marked with $\bigcirc$ are not quite consistent 
with each other within the respective statistical uncertainties. 
 
Similar comparisons are given for the multi-GeV region. 
The $\mu$-like "fully contained (FC) events" relate to all of 
the visible energy being contained in the inner detector and  
"partially contained (PC)" refers to events with some of 
the visible energy in the outer detector. 
For the identification procedure in the multi-GeV region, 
the Super K. group utilizes the characteristic 
\v{C}erenkov ring pattern of muons and electrons. 
A $\mu$-like event produces a clear outer ring edge of 
\v{C}erenkov light while an electron-like event creates 
a fuzzy ring pattern due to the production of cascade showers 
and multiple scattering of low-energy electrons~\cite{TKYT}. 
The comparison between the 33.0 kty and 46.3 kty cases are given 
in the lower part of Table 1. 
The intensities of $e$-like events marked with $\otimes$ appear to exhibit 
a large difference (of the order of \%) comparable to the sub-GeV case. 
Accordingly both intensities for ($e$-like events + $\mu$-like events) 
and also the ratio ($e$-like intesity)/($\mu$-like intesity), 
marked with $\ominus$ and $\odot$ respectively, 
are different beyond their statistical uncertainties. 
However, intensities for $\mu$-like (FC) and (PC) events 
for both exposures are consistent within the respective uncertainties.

\section{Study of the $e/\mu$ identification capability of the Super K. detector}

The only experimental test of the $e/\mu$ identification capability 
was carried out by part of the Super K. group using 
the 12-GeV proton synchrotron at KEK, Japan~\cite{SKa}. 
For the test, they used a 1000 ton water \v{C}erenkov detector. 
The water tank was cylindrical (about 10 m in height and 10 m in diameter) 
and its inner detector had a cylindrical volume of 
9.6 m in diameter and 9.3 m in height. 
It was equipped with 380 photomultiplier tubes (PMTs with 50-cm dia.). 
On the other hand the Super Kamiokande detector~\cite{YF} has 
a cylindrical volume, 39 m in diameter and 42 m in height, 
containing about 50\,000 metric tons of water. 
The fiducial volume of 32\,000 metric tons is cylindrical of 
33.8 m in diameter and 36.2m in height. 
It is equipped with 11\,146 50-cm PMTs 
(1773 PMTs each on the upper and lower surfaces and 
7600 PMTs on the  sides) which cover 40 \% of the inner surface. 
It is therefore conceivable that the large scale differences between 
the two detectors may lead a larger $e/\mu$  mis-identifications for 
the full Super K. detector than the one that was obtained for 
the 1000 ton exposures. 
In addition,  the experimental data from the test set-up are of 
low statistics and limited only to sub-GeV events. 
There are no results  for multi-GeV events because of 
the smallness of the 1000 ton detector.  
 
The Super K. group has performed a Monte Carlo (MC) calculation 
to establish a relation, $N_{MC} (\theta, P_{e})$, 
between the average number of photoelectrons (p.e.), 
the electron momentum $P_e$, and the angle of incidence $\theta$. 
The angle was between the particle direction 
and the center of circular area of 50 cm diameter. 
In the framework of this simulation it was assumed that 
the electron with a given momentum was started from a vertex 
at the center of sphere of 16.9 m radius. 
The relation $N_{MC} (\theta, P_{e})$ is based only on 
the average number of  p.e., neglecting the fluctuations around its average. 
By using this relation the expected p.e. number produced 
by the electron in the $i$-th PMT on the inner surface of 
the cylindrical volume was calculated with the following expression 
 
\[ 
N_{i,exp}(direct) = \alpha_e\times N_{MC}(\theta, P_{e})
\times \left(\frac{16.9\ {\rm m}}{l_i} \right)^{1.5} 
\times \exp \left(-\frac{l_i}{L} \right)
\times f(\Theta). 
\]   
 
This expression was given as equation (6.7) in 
Kasuga's PhD paper~\cite{SKas}. 
In this expression, the exponent of 1.5 for (16.9 m/$l_i$) 
should have been 2.0. 
Also a p.e. number distribution for muons is given in 
a similar expression (6.8), which is  
 
\begin{eqnarray} 
N_{i,exp}(direct) &=& 
\left(\alpha_\mu \times 
\frac{1}{l_i(\sin \theta_i + l_i \cdot(\frac{\rmd\theta}{\rmd x}))} 
\times \sin^2\theta_i + N_{i,knock}(\theta_i)
\right) 
\nonumber \\
&& 
\times \exp \left(-\frac{l_i}{L} \right) \times f(\Theta). 
\nonumber 
\end{eqnarray}
The explanation for the denominator in the second case is 
incorrect in Figure 6.6 of Kasuga~\cite{SKas} 
and Figure 2.14 of  Sakai~\cite{AS}, 
afterwards being corrected by Ishihara in Figure 5.7~\cite{KIsh}. 
Also they define the probability function  suitable for 
the Particle Identification (P.I.), given by 
 
\[ 
Prob(N_{exp}, N_{obs}) = 
\frac{1}{\sqrt{2\pi}\sigma} 
\exp\left(-\frac{(N_{obs} - N_{exp})^2}{2\sigma^2}\right), 
\]
with $\sigma^2$ = $1.2^2 \times N_{exp}$ + (0.1 $\times N_{exp})^2$. 
The first term with the factor 1.2 in $\sigma$ comes from 
the actual PMT charge resolution and the second factor takes into account 
the gain uncertainty in the PMT. 
The likelihood functions $L_e$ for $e$-like events and 
$L_\mu$ for $\mu$-like events shown in their expression (6.12) are 
obtained with the probability $Prob(N_{exp}, N_{obs})$ for studying 
the $e/\mu$ identification capability of the events, 
$L_e = \log (\Pi_{\theta_i<(1.5 \times \theta_c)}\, Prob_i(e))$ 
and $L_\mu = \log (\Pi_{\theta_i<(1.5 \times \theta_c)}\,Prob_i(\mu))$, 
with an opening angle of 1.5 $\times \theta_c$ from 
the reconstructed particle direction.
By adding information of the \v{C}erenkov opening angles 
combined with the information of the ring pattern, 
the used method is improved with a PID (Particle Identification) parameter 
as given in their expression (6.16) to obtain a reduced 
mis-identification probability.
Using the Monte Carlo simulation by applying these expressions to 
the experimental data obtained at the 12-GeV proton synchrotron, 
the group concluded~\cite{SKas, SK} that they were able to 
identify electrons and muons with mis-identification probabilities of 
less than a few \% in the momentum range 250-1000 MeV/$c$. 
In the similar study in Sakai's PhD paper~\cite{AS}, however, 
the mis-identifications for $\mu \to e$ are given as 
several \% and for $e \to \mu$ as a few \% 
as shown in Table 7.1 and Table 7.2 in his paper using 
the same expression with the exponent 2.0, 
without considering stochastic processes of cascade showers. 
The Super K. group does not apply the $e/\mu$ identification method 
for multi-GeV region. 
They argue that muons in multi-GeV range can produce 
very large p.e. numbers compared to electrons of the same momentum, 
even increasing for larger momenta. 
Therefore, the $e/\mu$ identification procedure applied 
for the sub-GeV range is unnecessary 
for multi-GeV events according to their estimates.      
  
\section{Our full MC simulation}

\begin{figure}[htbp]
\begin{center}
\includegraphics[width=10cm]{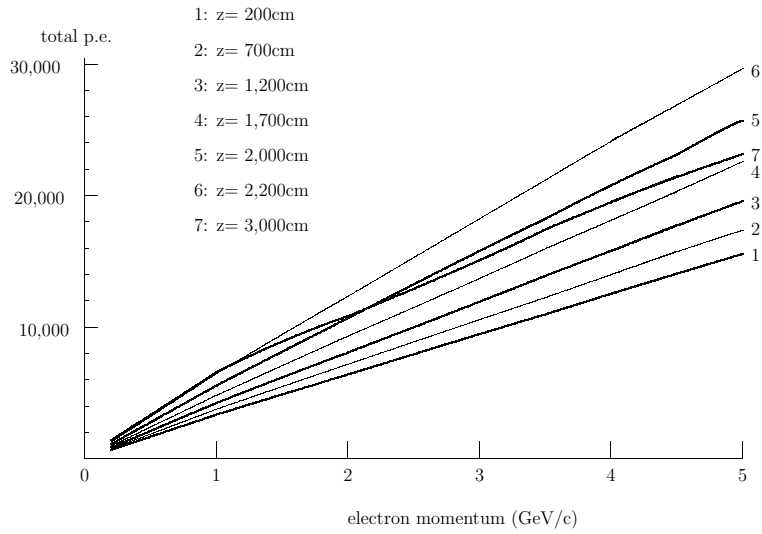}
\caption{Dependence of the average p.e. numbers on electron momenta in the sub-GeV and multi-GeV ranges at various z - values for vertical incidences.}
\label{fig:fige-p}
\end{center}
\end{figure}

\begin{figure}[htbp]
\begin{center}
\includegraphics[width=10cm]{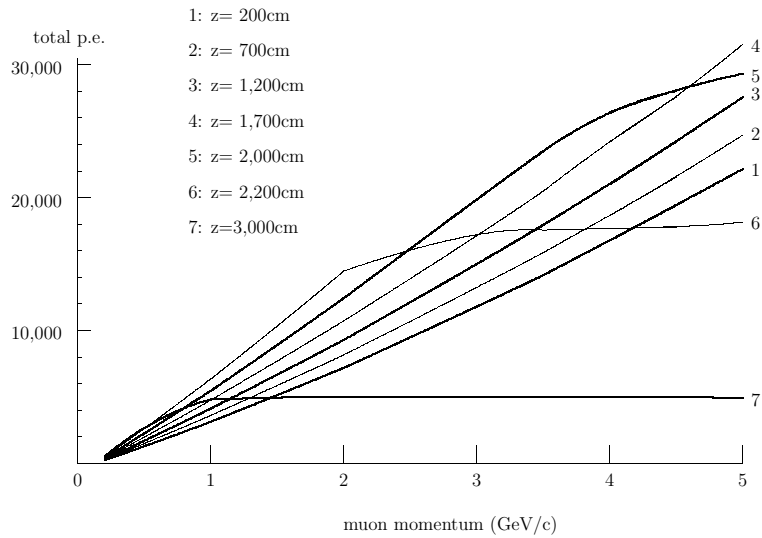}
\caption{Dependence of the average p.e. numbers on muon momenta in the sub- and multi-GeV ranges at various z - values for vertical incidences.}
\label{fig:figmu-p}
\end{center}
\end{figure}

The Super Kamiokande collaboration has determined 
the relations relevant for the mis-identification probabilities 
( $\theta_j, P_{e \; j}$ or $\theta_j, P_{\mu \; j}$) 
by only using the average number of photoelectrons. 
Of course, it is natural to expect that there are many events 
where the observed number of p.e. fluctuates around the average number. 
These fluctuations are not only caused by Poisson distributions, 
but also by the stochastic processes in 
the electromagnetic interactions of electrons and muons. 
Moreover, it is also an oversimplification to use 
only ionization and knock-on electron production for 
the expected p.e. numbers in the muon expression 
(equation (6.8) in the paper)~\cite{SKas}. 
However, in order to describe correctly the behaviors of fluctuations, 
additional energy-loss processes, 
like direct electron pair production and bremsstrahlung cannot be neglected. 
In the cylindrical fiducial volume, also, 
different path lengths of \v{C}erenkov light due to 
different vertex positions and incident angles contribute to 
the fluctuations of the p.e. numbers. 
Therefore we simulated 100 events each with different vertex positions 
and determined the average p.e. number, while the p.e. number obtained 
by the Kasuga expression (6.8) depends only on 
the distance between $l_i$ between the $i$-th PM and the vertex. 
To clearly demonstrate the effect we show 
the total photoelectron numbers for muons and electrons in the sub-GeV 
and multi-GeV ranges in their dependence on various vertex positions 
(different z-values) at x = y = 0 cm and for vertical incidence 
in \fref{fig:fige-p} and \ref{fig:figmu-p}.  


In view of such a complicated situation, 
we have performed a full MC simulation including 
all kinds of fluctuations using 
the EGS code for the Super Kamiokande detector. 
The EGS simulation can be initiated for electrons or photons 
which originate from a muon. 
As far as all other necessary quantities 
(water transparency, vertex position resolution and angular resolution etc.) 
are concerned, we follow the procedure used by the Super K. group. 
The step size of track lengths in our Monte Carlo is 
taken as 0.028 radiation lengths for electron events. 
Also we use the probability function defined by Kasuga~\cite{SKas} 
and determine the likelihood functions. 
As an appropriate coordinate system, 
we consider the x- and y-axes to be in a horizontal plane and 
the z-axis to extend perpendicularly downward. 
So the upper circular surface is at z= 0 cm and the lower one at z= 3400 cm. 
The non-linear behaviour in both figures reflect effects of 
vertex positions (corresponding to different z values), 
by which some tracks of events escape from the fiducial volume.

\begin{figure}[htbp]
\begin{center}
\includegraphics[scale=0.65]{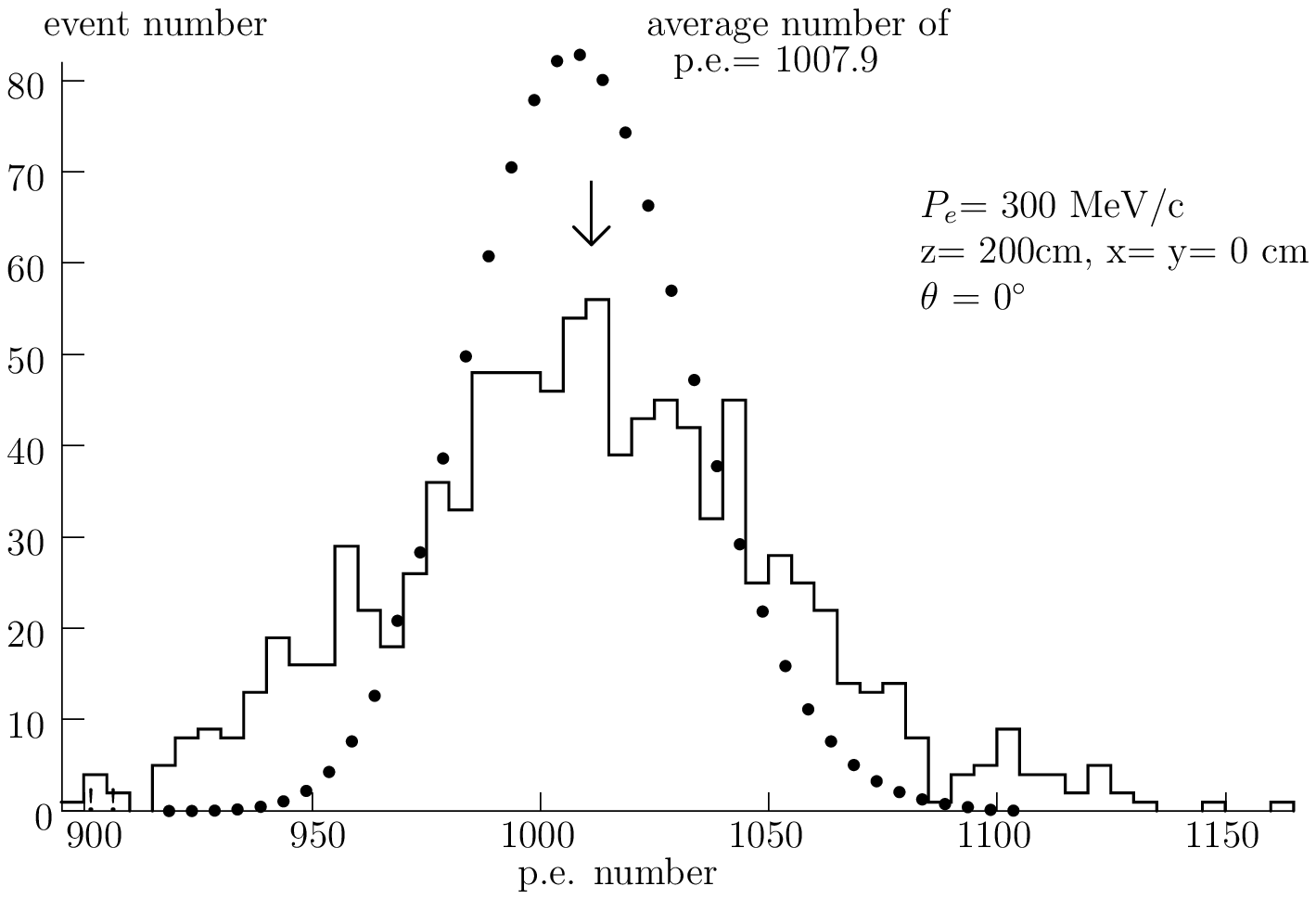}
\caption{Dependence of the p.e. number distribution for electrons of 300 MeV/$c$ with a vertex position at z=200cm.} 
\label{fig:pe3-2e}
\end{center}
\end{figure}
 
\begin{figure}[htbp]
\begin{center} 
\includegraphics[scale=0.65]{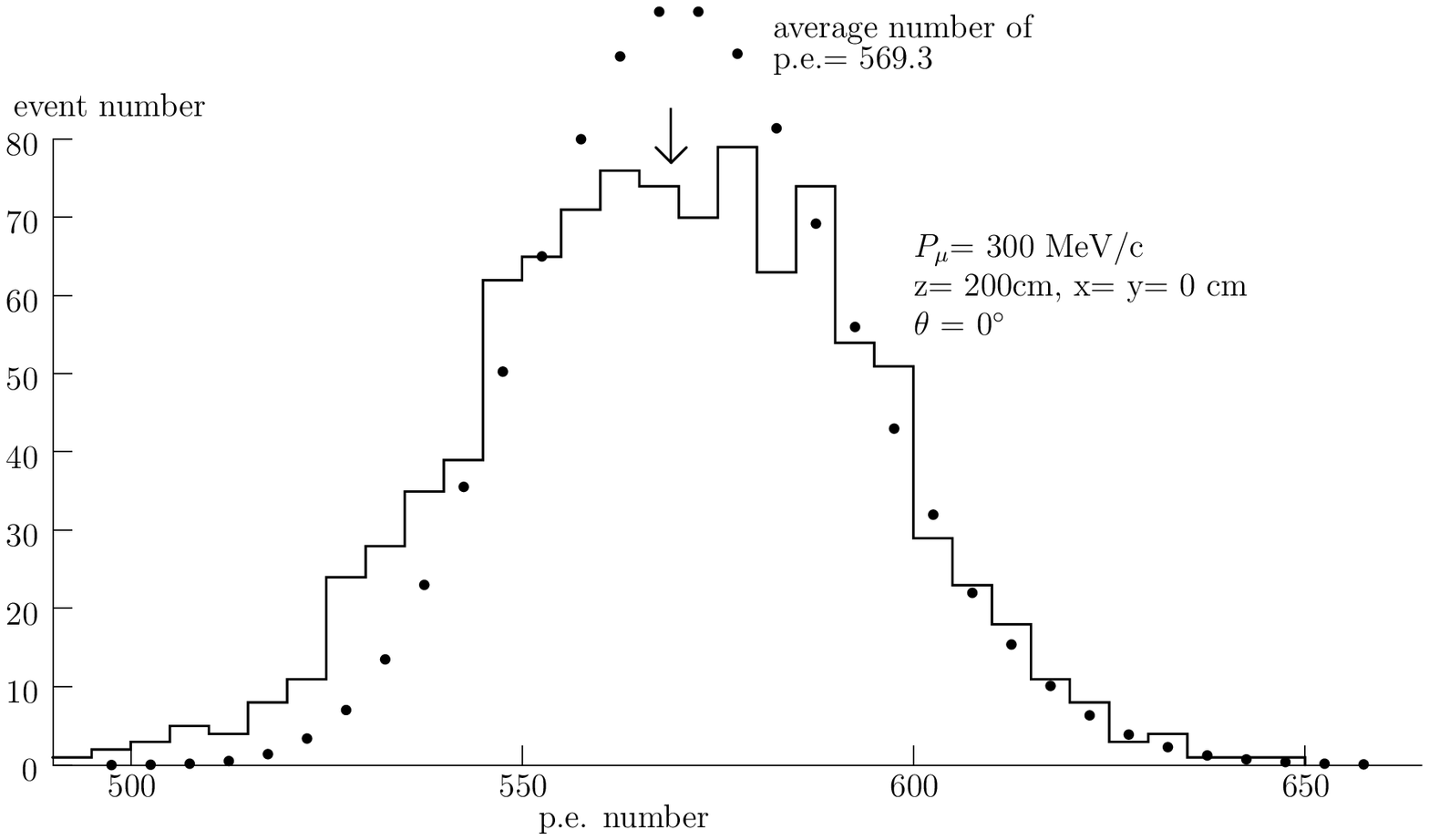}
\caption{Dependence of the p.e. number distribution for muons of 300 MeV/$c$ with a vertex position at z=200cm.}
\label{fig:pe3-2m}
\end{center}
\end{figure}
 
\begin{figure}[htbp]
\begin{center}
\includegraphics[scale=0.65]{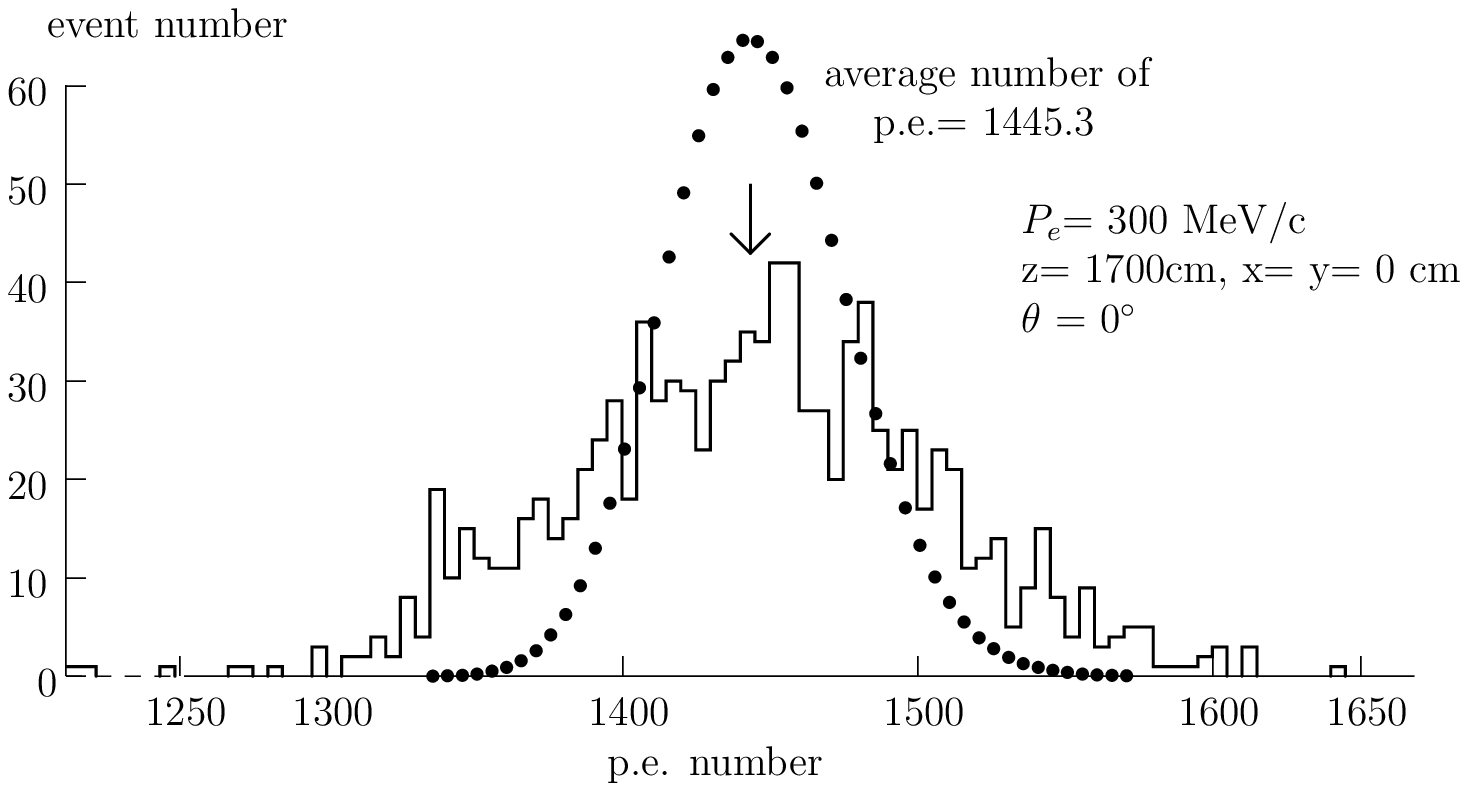}
\caption{Dependence of the p.e. number distribution for electrons of 300 MeV/$c$ with a vertex position at z=1700cm.}
\label{fig:pe3-17e}
\end{center}
\end{figure}
 
\begin{figure}[htbp]
\begin{center}
\includegraphics[scale=0.65]{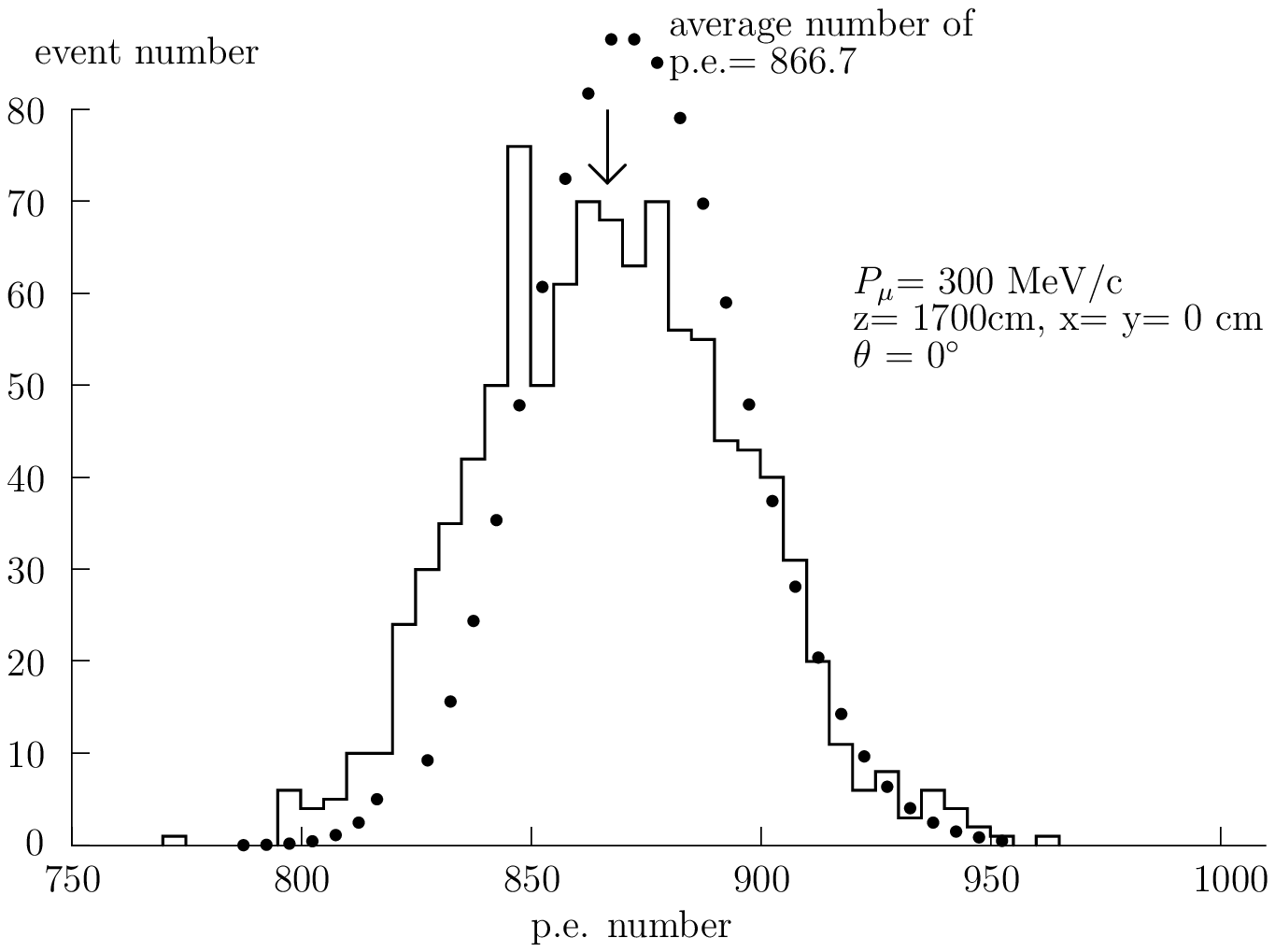}
\caption{Dependence of the p.e. number distribution for muons of 300 MeV/$c$ with a vertex position at z=1700cm.}
\label{fig:pe3-17m}
\end{center}
\end{figure}
 
\begin{figure}[htbp]
\begin{center}
\includegraphics[scale=0.65]{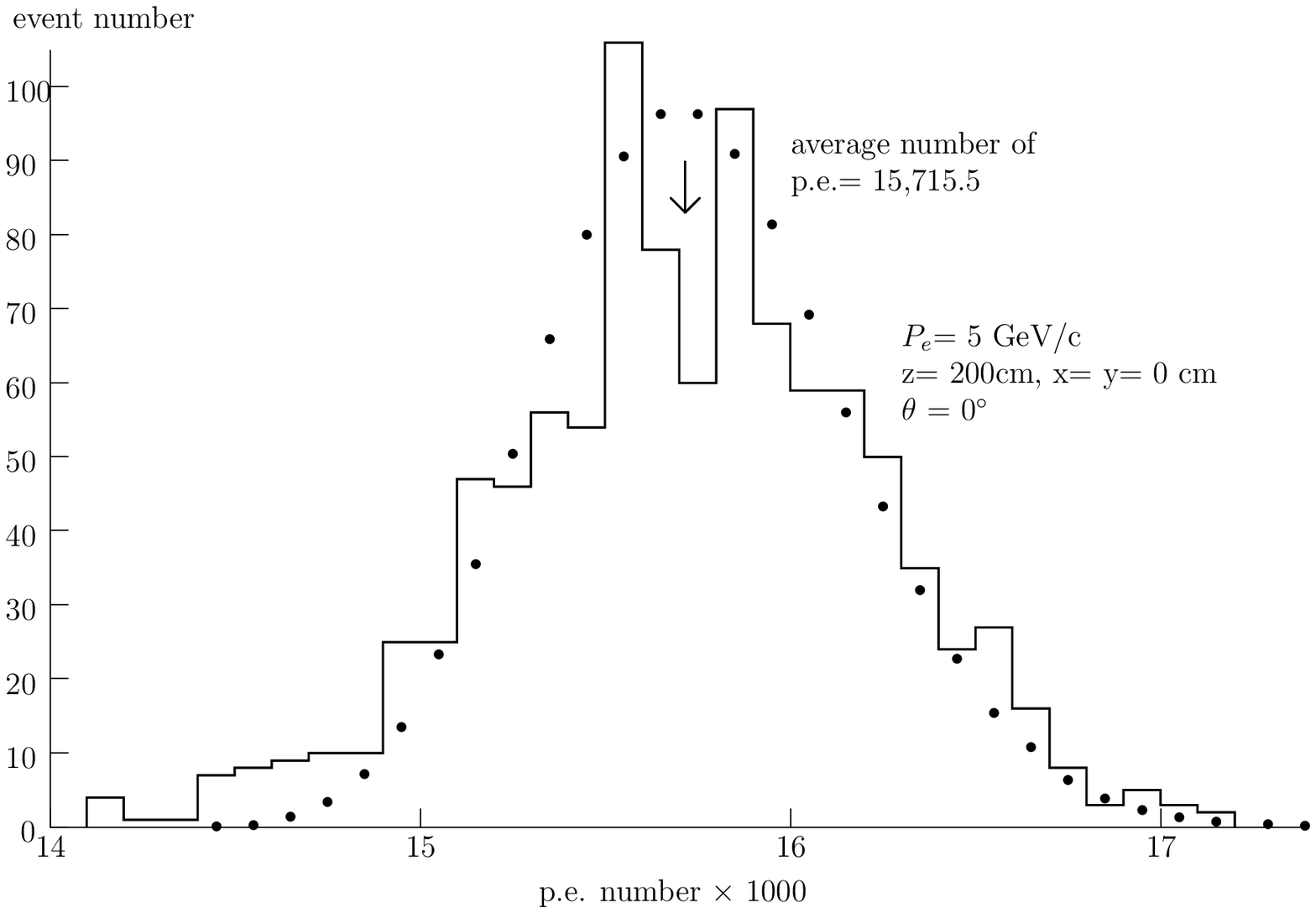}
\caption{Dependence of the p.e. number distribution for electrons of 5 GeV/$c$ with a vertex position at z=200cm.}
\label{fig:pe50-2e}
\end{center}
\end{figure}
 
\begin{figure}[htbp]
\begin{center}
\includegraphics[scale=0.65]{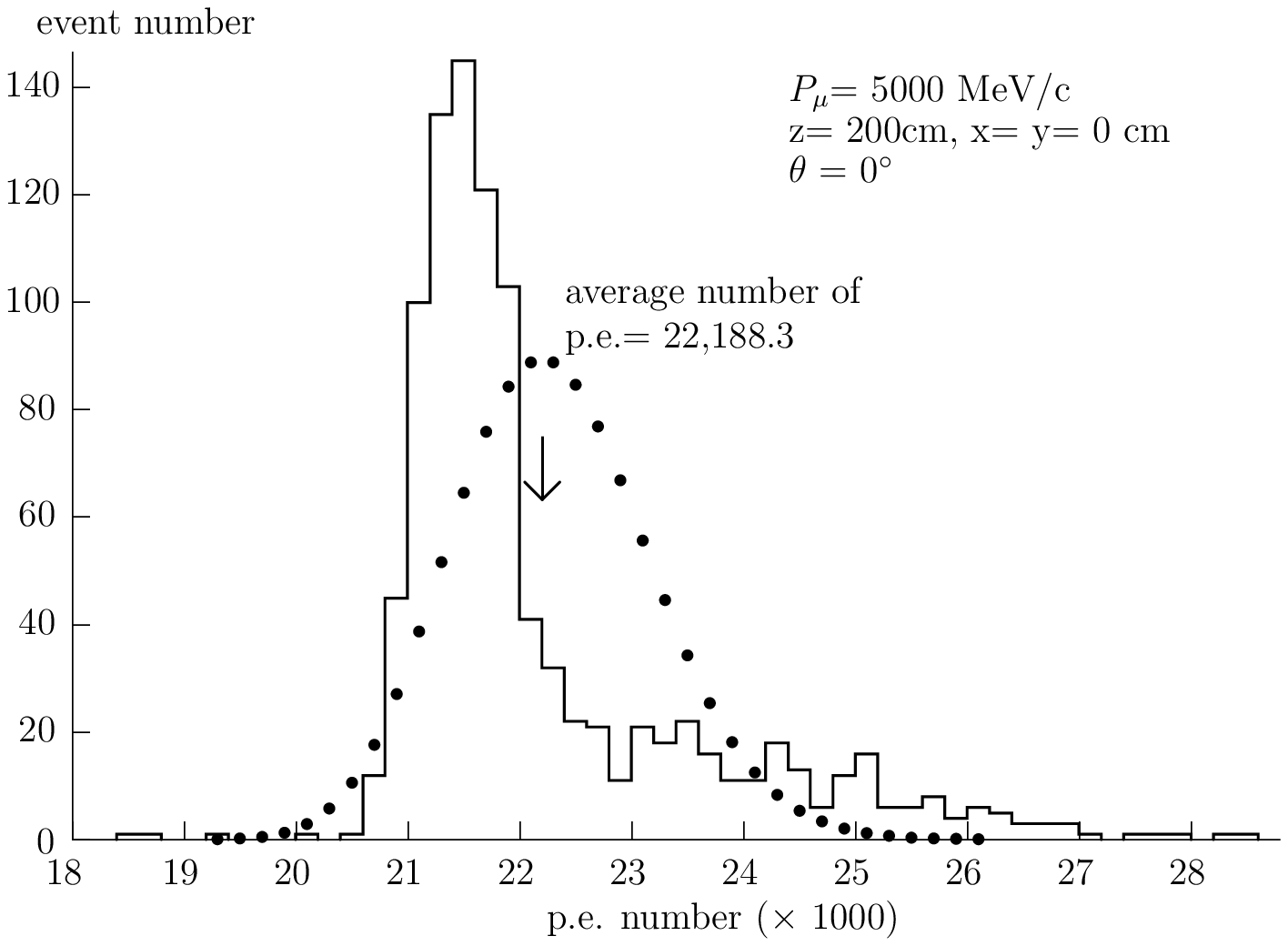}
\caption{Dependence of the p.e. number distribution for muons of 5 GeV/$c$ with a vertex position at z=200cm.}
\label{fig:pe50-2m}
\end{center}
\end{figure}
 
\begin{figure}[htbp]
\begin{center}
\includegraphics[scale=0.65]{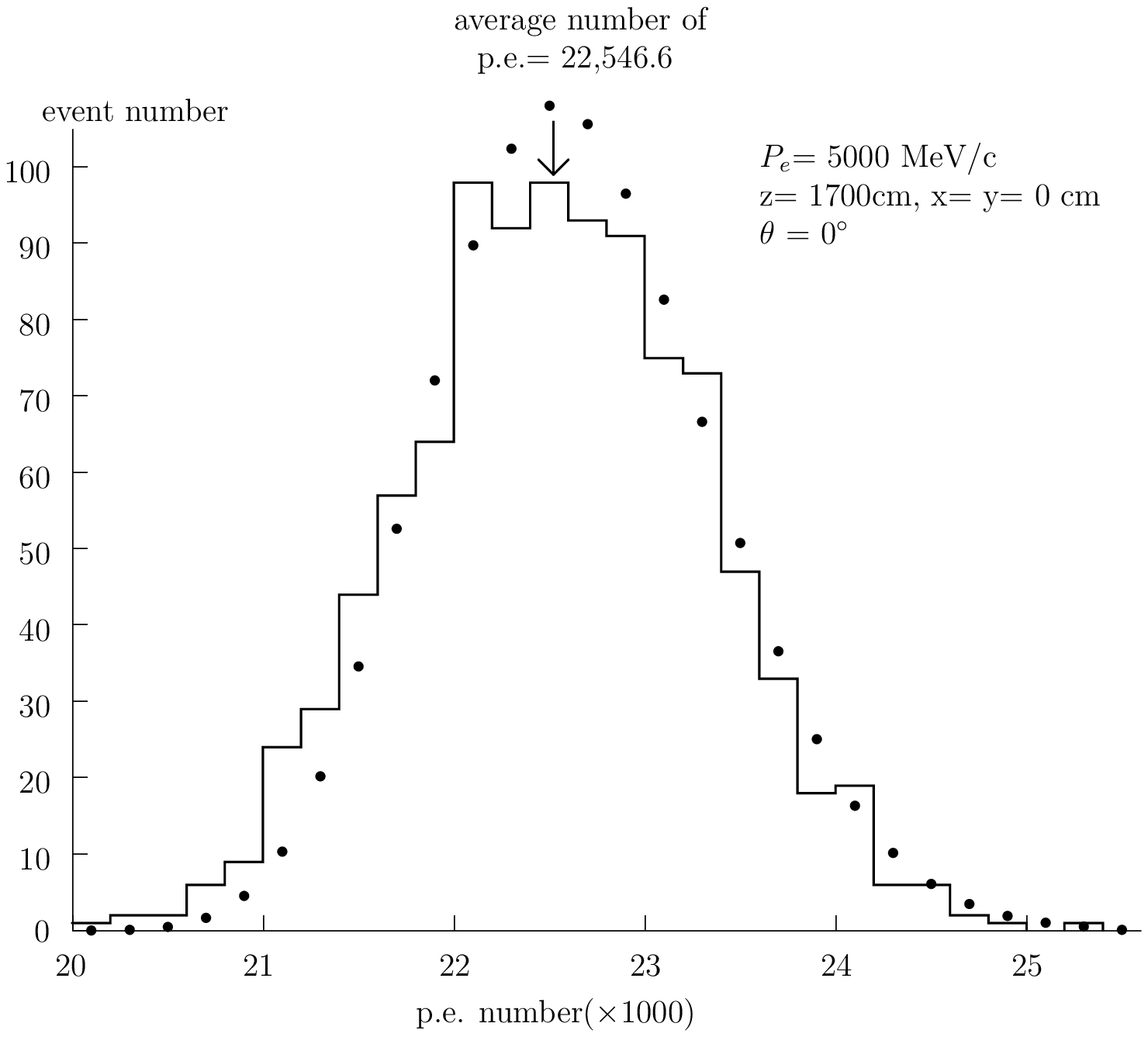}
\caption{Dependence of the p.e. number distribution for electrons of 5 GeV/$c$ with a vertex position at z=1700cm.}
\label{fig:pe50-17e}
\end{center}
\end{figure}
 
\begin{figure}[htbp]
\begin{center}
\includegraphics[scale=0.65]{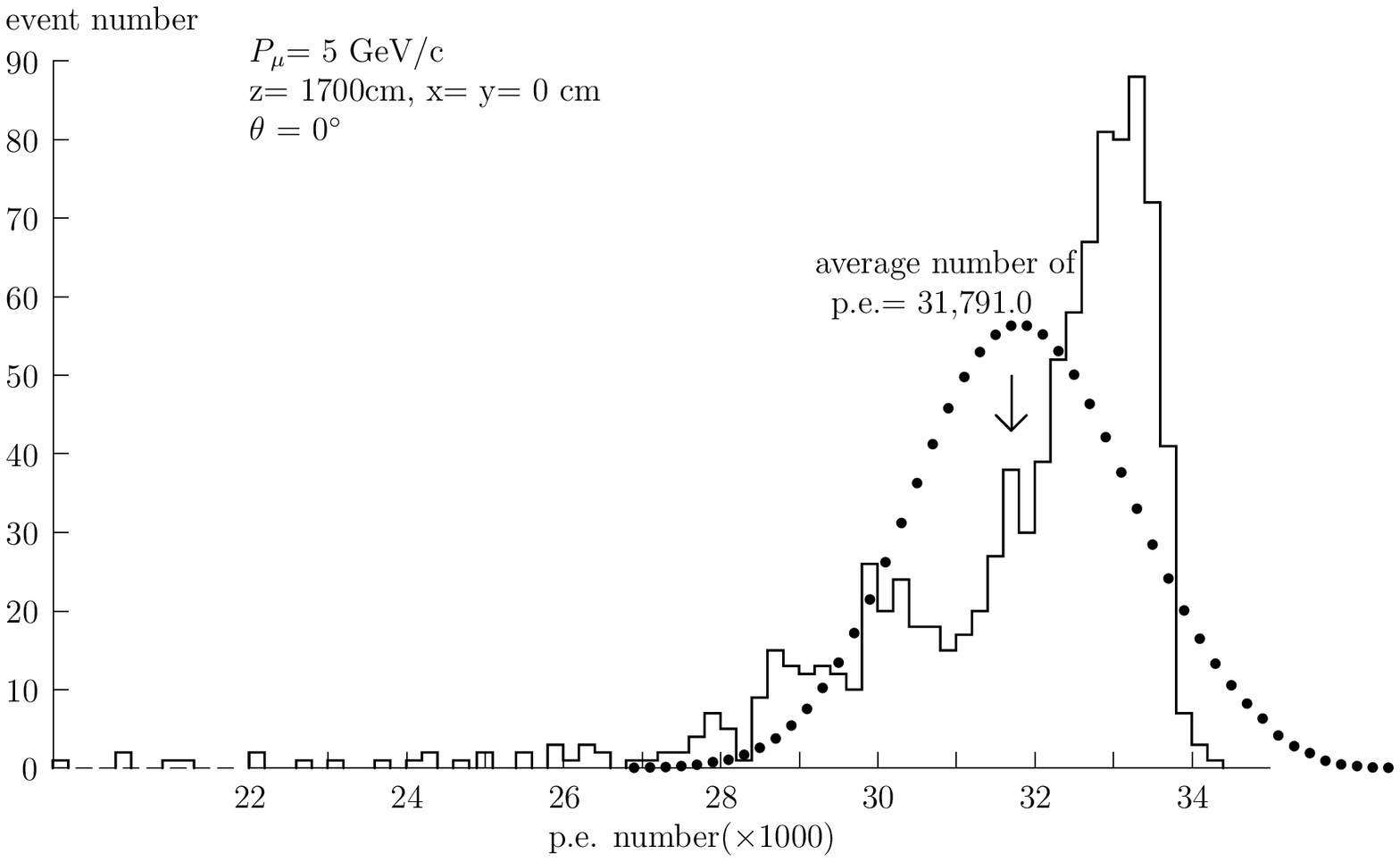}
\caption{Dependence of the p.e. number distribution for muons of 5 GeV/$c$ with a vertex position at z=1700cm.}
\label{fig:pe50-17m}
\end{center}
\end{figure}

As a reference, figures \ref{fig:pe3-2e} and \ref{fig:pe3-2m} show 
the fluctuation pattern of electrons and of muons of 300 MeV/$c$ for 
the vertex position of z= 200 cm and $\theta = 0^\circ$ 
obtained with 1000 MC events. 
The average p.e. number for electrons with 300 MeV/$c$ is 
larger than that of muons with the same momentum, 
but the situation is reversed at 5 GeV/$c$. 
The full black points in the respective figures show 
the Poisson distribution calculated for each average p.e. number 
where the track started at the respective vertex position. 
Figures \ref{fig:pe3-17e} and \ref{fig:pe3-17m} give the results for 
the same momentum at a different vertex point (z = 1700 cm). 
Figures \ref{fig:pe50-2e} and 
\ref{fig:pe50-2m} are for 5 GeV/$c$ at z = 200 cm 
and figures \ref{fig:pe50-17e} and \ref{fig:pe50-17m} 
for 5 GeV/$c$ at z = 1700 cm. 
In these figures we have shown only events for 
the minimum momentum value in sub-GeV range and 
the maximum momentum value in multi-GeV range. 
These results show that electrons of 300 MeV/$c$ started at 
z = 200 cm and 1700 cm show comparatively larger deviations from 
the Poisson distributions than those of muons under the same condition. 
On the other hand muons of 5 GeV/$c$ at both given z-values exhibit 
larger fluctuations than electrons. 
Also, for the inclined incidence directions similar fluctuations occur 
with even stronger deviations compared to the vertical case. 
Accordingly, the fully simulated MC results clearly prove that 
the fluctuations owing to the stochastic processes can produce 
a serious impact on the identification problems of muon and electron events 
not only in the sub-GeV region but also in the multi-GeV region 
through their effect on the ring patterns. 
   
\subsection{Our $L_{\mu}-L_e$ distributions}

We inspected the $L_{\mu}-L_e$ distributions for electron 
and muon events with various momenta in the sub-GeV range 
and also some in the multi-GeV range 
for various starting positions (z-values), keeping x = y = 0. 
Then, the momenta of electrons and muons were selected 
such as to produce almost the same p.e. numbers. 
The mis-identification rates between both event types are obtained 
from their estimated results as obtained from 
the $L_{\mu}-L_e$ distributions. 
For this purpose 500 muon and electron events each were used. 
As shown in table \ref{table:lule1}, 
the corresponding mis-identification probabilities 
are summarized and identified by various symbols. 
The mis-identification rates for muons (\% values) 
as explained below the table are defined in the following way: 
[(muon numbers misidentified as electrons) ${}-{}$ 
(electron numbers misidentified as muons)]/(original number of muons). 
The mis-identification rates appear to be larger than 
the ones given in Kasuga's paper. 
From the results shown in the table, one might consider it difficult 
to reliably distinguish electron from muon events 
in particular in the sub GeV region, 
although the situation is improving with increasing momentum. 
However, these mis-identification rates are only valid 
for the case of x = y = 0. 
 
\begin{table}[htbp]
\begin{center}
\caption{$L_{\mu}$--$L_{e}$ distributions between electrons and muons in sub-GeV and some of multi-GeV.}
\label{table:lule1}
\small
\begin{tabular}{|c|c|c|c|c|c|}\hline 
{\bf z(cm)} & 
$100e/225\mu$& $200e/310\mu$ & $300e/400\mu$ & $400e/500\mu$ & $500e/560\mu$ \\
\hline 
{\bf 200}&
$\times$& $\times$& $\times$& $\triangleleft$ & $\triangleleft$ \\ \hline 
{\bf 700}&
$\times$& $\times$& $\times$&$\times$& $\times$\\ \hline 
{\bf 1200}&
$\bigcirc *$& $\times$& $\times$&$\times$& $\times$\\ \hline 
{\bf 1700}&
$\bigcirc *$& $\times$& $\triangleleft$ &$\bigtriangleup$&$\bigtriangleup$ \\ 
\hline 
{\bf 2200}&
$\bigcirc *$& $\circ$&$\bigtriangleup$&$\bigtriangleup$&$\bigtriangleup$ \\ 
\hline 
{\bf 2700}&
$\times$& $\circ$&$\bigtriangleup$&$\times$&$\triangleleft$ \\ 
\hline 
{\bf 3000}&
$\triangleleft$& $\bigtriangleup$&$\triangleleft$& $\times$& $\times$ \\ 
\hline \hline
{\bf z(cm)} & 
$600e/650\mu$& $800e/820\mu$ & $1000e/1000\mu$ & 
$2000e/2000\mu$ & $3000e/3000\mu$ \\
\hline 
{\bf 200}&
$\triangleleft$ & $\circ$& $\circ$&$\bigcirc$ &$\bigcirc$ \\ \hline 
{\bf 700}&
$\times$& $\bigcirc$ & $\circ$&$\bigcirc$&$\bigcirc$ \\ \hline 
{\bf 1200}&
$\triangleleft$& $\bigcirc$& $\circ$&$\bigcirc$& $\bigcirc$\\ \hline 
{\bf 1700}&
$\circ$& $\circ$& $\circ$ &$\bigcirc$&$\bigcirc$ \\ 
\hline 
{\bf 2200}&
$\bigtriangleup$& $\circ$&$\circ$&$\bigcirc$&$\bigcirc$ \\ 
\hline 
{\bf 2700}&
$\bigtriangleup$&$\bigtriangleup$ &$\bigtriangleup$&$\circ$&$\circ$ \\ 
\hline 
{\bf 3000}&
$\times$& $\times$&$\times$& $\times$& $\triangleleft$\\ 
\hline 
\multicolumn{6}{c}{}\\
\multicolumn{6}{l}{{\bf the resultant mis-identification rate for muons;}}\\
\multicolumn{6}{l}{$\bigcirc$: $0\sim5$ \%, $\circ$: $6\sim20$ \%,
$\bigtriangleup$: $21\sim60$ \%, $\triangleleft$: $61\sim95$ \%,
$\times$: $96\sim100$ \%}\\
\multicolumn{6}{l}{$*$: means $L_{\mu}$--$L_{e}$ distribution with the opposite signs.}
\end{tabular}
\normalsize
\end{center}
\end{table}
 
\begin{figure}[htbp]
\begin{center}
\includegraphics[width=10cm]{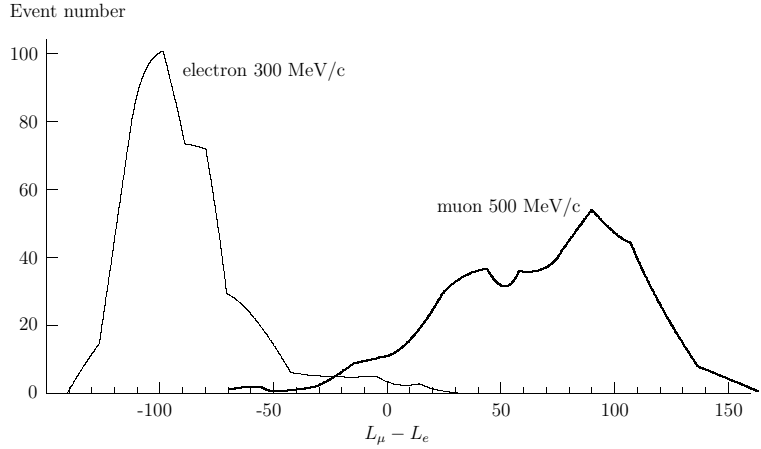}
\caption{$L_{\mu}- L_e$ for 300 MeV/$c$ electrons and for 500 MeV/$c$ muons at a starting point of x = y = 0 and z = 2,700 cm.}
\label{fig:zu3-5}
\end{center}
\end{figure}

\begin{figure}[htbp]
\begin{center}
\includegraphics[width=10cm]{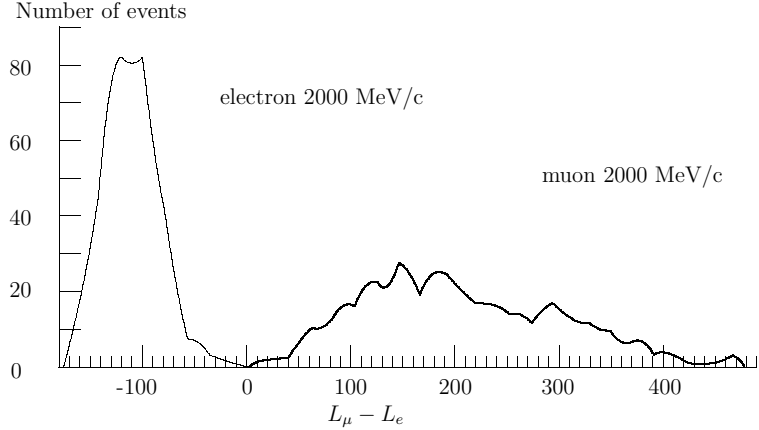}
\caption{$L_{\mu}- L_e$ for 2000 MeV/$c$ electrons and for 2000 MeV/$c$ muons at a starting point of x = y = 0 and z = 1,200 cm.}
\label{fig:figure3}
\end{center}
\end{figure}

As an example, we show in figure \ref{fig:zu3-5} 
the distributions of $L_{\mu}-L_e$ for 300 MeV/$c$ electrons and 
for 500 MeV/$c$ muons 
(the events with the chosen momenta will radiate 
almost similar p.e. numbers on average) for z = 2700 cm. 
In some cases it is difficult to distinguish betweem the two event types. 
Figure \ref{fig:figure3} gives the same quantity 
for 2000 MeV/$c$ electrons and for 2000 MeV/$c$ muons 
at z = 1200 cm which allows to distinguish between the two event types. 
But, this is only true for vertical incidence keeping x = y = 0. 
One should notice that the $L_{\mu}-L_e$ distribution for muons is 
much wider than that for electrons. 
This aspect is very likely caused by stronger fluctuations 
due to stochastic interaction processes of muons rather than 
fluctuations due to the shower development of electrons. 
As seen in table \ref{table:lule1}, 
the $L_{\mu}-L_e$ distributions of electrons and muons 
would lead to a mis-identification of $\le$ few \% 
for muons in the region of z = 200 cm $\sim$ 2200 cm typically 
in the multi-GeV region. 
However, these results are only true for events started 
at the given vertex position of x = y = 0 
and for vertical incidence. 
For z \textgreater 2200 cm, 
mis-identification rates of $\textgreater$ several \% for muons are obtained, 
and it has to be kept in mind that 
some muons could escape through the bottom surface of the detector. 
Moreover, such muons with oblique incidences can generate 
$L_{\mu}-L_e$ distributions differently shaped 
compared to those for vertical incidence.  

\subsection{Our results with PID parameter}

Following Kasuga's method of Particle Identification (PID) 
using the pattern and opening angle of the \v{C}erenkov ring, 
furthermore, we calculated the PID parameter values. 
For the sub-GeV region, 
we obtained the \% values for muon deficits in table \ref{table:table2a} 
using a number of 9240 for electrons and 14\,910 for muons 
starting at vertex points from z = 200 cm to 3000 cm 
with their incidence direction $\theta = 0^\circ$ along the z-axes 
(x = 0 cm, and y = 0 cm, 50 cm, 100 cm, 250 cm, 500 cm and 750 cm). 
The event numbers are taken from the intensities of the $\nu_e$ 
and $\nu_{\mu}$ energy spectra~\cite{MH}. 
Then, according to our simulation, 4312 electrons and 6958 muons 
are estimated to stop up to z = 1400 cm. 
$e \to$ $\mu$ means the electron is misidentified as a  muon, 
and $\mu \to e$ the opposite. 
The table shows substantial shifts from muons to electrons. 
 
\begin{table}[htbp]
\begin{center}
\caption{Summarized PID results for distributions between electron and muon events in the sub-GeV region.} 
\label{table:table2a}
\footnotesize
\begin{tabular}{|c||r|r|r|r||r|r|r|r||}\hline 
\Gcenter{2}{Sub-GeV} & 
\multicolumn{2}{|l|}{{\bf z=200 $\sim {}$}} & 
\multicolumn{2}{|l||}{{\bf z=200 $\sim {}$}} &
\multicolumn{2}{|l|}{{\bf z=200 $\sim {}$}} &
\multicolumn{2}{|l||}{{\bf z=200 $\sim {}$}} \\ 
\Gcenter{2}{region}& \multicolumn{2}{|c|}{{\bf 1400 cm}} &
\multicolumn{2}{|c||}{{\bf 3000 cm}} &
\multicolumn{2}{|c|}{{\bf 1400 cm}} &
\multicolumn{2}{|c||}{{\bf 3000 cm}} \\ \cline{2-9}
&$e$&$\mu$&$e$&$\mu$&$e$&$\mu$&$e$&$\mu$ \\ \hline \hline
&\multicolumn{4}{|l||}{{\bf 1) x=0, y=0 cm}} &
\multicolumn{4}{|l||}{{\bf 2) x=0, y=50 cm}} \\ \hline
event number & 4312 & 6958 & 9240 & 14\,910 & 
4312 & 6958 & 9240 & 14\,910 \\ \hline
$e \rightarrow \mu$, $\mu \rightarrow e$ & 
1 & 641 & 143 & 1202 & 78 & 3153 & 611 & 4154 \\ \hline
increase & +14.8 \% & $-9.2$ \% & +11.5 \% & $-7.1$ \% &
+71.3 \% & $-44.1$ \% & +38.3 \% & $-23.8$ \% \\ \hline \hline
& \multicolumn{4}{|l||}{{\bf 3) x=0, y=100 cm}} &
\multicolumn{4}{|l||}{{\bf 4) x=0, y=250 cm}} \\ \hline 
event number & 4312 & 6958 & 9240 & 14\,910 & 
4312 & 6958 & 9240 & 14\,910 \\ \hline
$e \rightarrow \mu$, $\mu \rightarrow e$ & 
75 & 3334 & 672 & 4411 & 98 & 3479 & 798 & 4615 \\ \hline
increase & +75.6 \% & $-46.8$ \% & +40.5 \% & $-25.1$ \% &
+78.4 \% & $-48.6$ \% & +41.4 \% & $-25.6$ \% \\ \hline \hline
& \multicolumn{4}{|l||}{{\bf 5) x=0, y=500 cm}} &
\multicolumn{4}{|l||}{{\bf 6) x=0, y=750 cm}} \\ \hline 
event number & 4312 & 6958 & 9240 & 14\,910 & 
4312 & 6958 & 9240 & 14\,910 \\ \hline
$e \rightarrow \mu$, $\mu \rightarrow e$ & 
140 & 3541 & 846 & 3924 & 106 & 3485 & 688 & 4735 \\ \hline
increase & +78.9 \% & $-48.9$ \% & +33.0 \% & $-20.6$ \% &
+78.4 \% & $-48.6$ \% & +43.8 \% & $-27.1$ \% \\ \hline \hline
& \multicolumn{4}{|l||}{{\bf 7) x=0, y=1000 cm}} &
\multicolumn{4}{|l||}{{\bf 8) x=y=0 cm, $\theta=10^{\circ}$}} \\ \hline 
event number & 4312 & 6958 & 9240 & 14910 & 
4312 & 6958 & 9240 & 14\,910 \\ \hline
$e \rightarrow \mu$, $\mu \rightarrow e$ & 
25 & 3307 & 596 & 4570 & 6 & 3995 & 211 & 5694 \\ \hline
increase & +76.1 \% & $-47.2$ \% & +43.0 \% & $-26.7$ \% &
+92.5 \% & $-57.3$ \% & +59.3 \% & $-36.8$ \% \\ \hline \hline
& \multicolumn{4}{|l||}{{\bf 9) x=y=0 cm, $\theta=20^{\circ}$}} &
\multicolumn{4}{|l||}{{\bf 10) x=y=0 cm, $\theta=30^{\circ}$}} \\ \hline 
event number & 4312 & 6958 & 9240 & 14\,910 & 
4312 & 6958 & 9240 & 14\,910 \\ \hline
$e \rightarrow \mu$, $\mu \rightarrow e$ & 
3 & 4197 & 173 & 6421 & 1 & 4226 & 142 & 6877 \\ \hline
increase & +97.3 \% & $-60.3$ \% & +67.6 \% & $-41.9$ \% &
+98.0 \% & $-60.7$ \% & +72.9 \% & $-45.2$ \% \\ \hline 
\end{tabular}
\normalsize
\end{center}   
\end{table}
 
For instance, in case of both events starting at z = 200 $\sim$ 3000 cm 
and at x = y = 0 with $\theta$ = $0^\circ$, 
the percentage of misidentified muons is 
$-7.1$ \% = $(143-1202)/14\,910$ and 
$+11.5$ \% = $(-143+1202)/9240$ for electrons. 
From these values, we obtain 
(number of muons misidentified as electrons) ${}-{}$ 
(number of electrons misidentified as muons) 
a deficit of 7.1 \% of muons and an 11.5 \% increase of electrons, 
i.e. the interactions of muon neutrinos would be 
systematically misinterpreted as those of electron neutrinos. 
For reference, some results for inclined incidence 
($\theta = 10^\circ$, $20^\circ$ and  $30^\circ$) 
at vertex positions of x = y = 0 cm are added. 
Integrating numerically these values for muon deficits considering 
the number of neutrinos as given 
by the $\nu_e$ ($\nu_{\mu}$) energy spectrum, 
we obtain a total mis-identification between electron and muon events 
for starting positions at z = 200 $\sim$ 3000 cm to be 
at least $\ge$ 20 \% for the sub-GeV region. 
 
\begin{table}[htbp]
\begin{center}
\caption{Summarized PID results for distributions between electron and muon events in the multi-GeV region.} 
\label{table:table2b}
\footnotesize
\begin{tabular}{|c||r|r|r|r||r|r|r|r||}\hline 
\Gcenter{2}{Multi-GeV} & 
\multicolumn{2}{|l|}{{\bf z=200 $\sim {}$}} & 
\multicolumn{2}{|l||}{{\bf z=200 $\sim {}$}} &
\multicolumn{2}{|l|}{{\bf z=200 $\sim {}$}} &
\multicolumn{2}{|l||}{{\bf z=200 $\sim {}$}} \\ 
\Gcenter{2}{region}& \multicolumn{2}{|c|}{{\bf 1400 cm}} &
\multicolumn{2}{|c||}{{\bf 2900 cm}} &
\multicolumn{2}{|c|}{{\bf 1400 cm}} &
\multicolumn{2}{|c||}{{\bf 2900 cm}} \\ \cline{2-9}
&$e$&$\mu$&$e$&$\mu$&$e$&$\mu$&$e$&$\mu$ \\ \hline \hline
&\multicolumn{4}{|l||}{{\bf 1) x=0, y=0 cm}} &
\multicolumn{4}{|l||}{{\bf 2) x=0, y=100 cm}} \\ \hline
event number & 949 & 1190 & 1897 & 2378 & 
949 & 1190 & 1897 & 2378 \\ \hline
$e \rightarrow \mu$, $\mu \rightarrow e$ & 
5 & 18 & 95 & 167 & 23 & 37 & 88 & 188 \\ \hline
increase & +1.4 \% & $-1.1$ \% & +3.8 \% & $-3.0$ \% &
+1.5 \% & $-1.2$ \% & +5.3 \% & $-4.2$ \% \\ \hline \hline
&\multicolumn{4}{|l||}{{\bf 3) x=0, y=250 cm}} &
\multicolumn{4}{|l||}{{\bf 4) x=0, y=500 cm}} \\ \hline
event number & 949 & 1190 & 1897 & 2378 & 
949 & 1190 & 1897 & 2378 \\ \hline
$e \rightarrow \mu$, $\mu \rightarrow e$ & 
4 & 41 & 81 & 73 & 14 & 57 & 88 & 236 \\ \hline
increase & +3.9 \% & $-3.1$ \% & $-0.4$ \% & +3.0 \% &
+4.5 \% & $-3.6$ \% & +7.8 \% & $-6.2$ \% \\ \hline \hline
&\multicolumn{4}{|l||}{{\bf 5) x=0, y=750 cm}} &
\multicolumn{4}{|l||}{{\bf 6) x=y=0 cm, $\theta=10^{\circ}$}} \\ \hline
event number & 949 & 1190 & 1897 & 2378 & 
949 & 1190 & 1897 & 2378 \\ \hline
$e \rightarrow \mu$, $\mu \rightarrow e$ & 
18 & 118 & 102 & 309 & 7 & 13 & 84 & 58 \\ \hline
increase & +10.5 \% & $-8.4$ \% & +10.9 \% & $-8.7$ \% &
+0.6 \% & $-0.5$ \% & $-1.4$ \% & +1.1 \% \\ \hline \hline
&\multicolumn{4}{|l||}{{\bf 7) x=y=0 cm, $\theta=20^{\circ}$}} &
\multicolumn{4}{|l||}{{\bf 8) x=y=0 cm, $\theta=30^{\circ}$}} \\ \hline
event number & 949 & 1190 & 1897 & 2378 & 
949 & 1190 & 1897 & 2378 \\ \hline
$e \rightarrow \mu$, $\mu \rightarrow e$ & 
31 & 14 & 148 & 38 & 5 & 41 & 81 & 73 \\ \hline
increase & +1.8 \% & $-1.4$ \% & $-5.8$ \% & +4.6 \% &
+3.8 \% & $-3.0$ \% & $-0.4$ \% & +0.3 \% \\ \hline 
\end{tabular}
\normalsize
\end{center} 
\end{table}
 
In the multi-GeV region, 
we used 1897 Monte Carlo events for electrons 
and 2380 MC events for muons as predicted by the respective neutrino fluxes. 
Similar PID parameters were calculated with vertex positions of 
z = 200 cm to 1400 cm and of z = 200 cm to 2900 cm. 
Some of the muons started at z $\textgreater$ 1400 cm 
would escape from the fiducial volume. 
Table \ref{table:table2b} again indicates the deficits of muons 
similar as in the sub-GeV region. 
This leads us, thus, to conclude that even in the multi-GeV region 
the analysis procedure used by the Super Kamiokande group 
shifts about more than several \% of muons into the electron category.

For checking with the ring patterns, 
we inspected 300 events each randomly selected from 1897 electron events 
and 2380 muon events starting at vertex positions from 
z = 200 cm to 1400 cm for vertical incidence along 
the z-axis (x = 0 cm, and y = 0 cm, 100 cm, 250 cm, 500 cm and 750 cm). 
This resulted in a number of cases where it was difficult 
to distinguish muon from electron events 
which corresponded to 5 \% and 11 \%, respectively, in each 300 event sample.
 
\section{Comparisons of the zenith angle distributions in the sub-GeV and multi-GeV ranges}

Subdivided into the 33 kty (the first half) 
and 46.3 kty (the last half) expressions from the total 79.3 kty exposure, 
we compare both the observed zenith angle distributions 
for the sub-GeV and multi-GeV ranges using $\chi^2$-tests~\cite{WHP}. 
All the numbers of events can be read off the Super K. papers. 
Although the Super K. group uses 10 bins 
for the observed zenith angle distribution of the complete 79.3 kty 
exposure~\cite{JKre}, the 10 bins are only available for 
the 79.3 kty exposure for the sub-GeV and multi-GeV ranges. 
Therefore, we can not inspect the distributions of 
33 kty and 46.3 kty exposures with 10 bins. 
Accordingly, we subdivided the whole observed zenith angle distributions 
separately into 5 bins for the respective distributions of 
33 kty and 46.3 kty exposures. 
In table \ref{table:chitab}, 
their zenith angle distributions of various events are represented 
with 5 bins for $-1 \le \cos \theta \le 1$. 
Further, for comparing both the upward-directed 
and downward-directed events among them, 
we subdivided the data into 3 bins each for $-1 \le \cos \theta \le 0$ 
and $0 \le \cos \theta \le 1$. 
The region of $-0.2 \le \cos \theta \le 0.2$ being near 
the horizontal direction was included for both distributions. 
 
\begin{table}[htbp]
\begin{center}
\caption{Chi-square tests for comparisons of the zenith angle distributions between 33 kty and 46.3 kty exposures of $e$-like and $\mu$-like events in the sub-GeV and the multi-GeV regions.}
\label{table:chitab}
\small
\begin{tabular}{||r@{.}l@{ $\sim$ }r@{.}l||r|r||r|r||r|r||} \hline
\multicolumn{4}{||c||}{zenith angle}&
\multicolumn{2}{|c||}{sub-GeV $e$-like}&
\multicolumn{2}{|c||}{sub-GeV $\mu$-like}&
\multicolumn{2}{|c||}{multi-GeV $e$-like} \\ \hline
\multicolumn{4}{||c||}{$\cos \theta$}&
33 kty & 46.3 kty & 33 kty & 46.3 kty & 33 kty & 46.3 kty \\ \hline
$-1$&0&$-0$&6& 288&349&183&291&50&41 \\ \hline
$-0$&6&$-0$&2& 233&344&226&273&55&73 \\ \hline
$-0$&2&0&2& 258&306&229&304&70&95 \\ \hline
0&2&0&6& 226&304&265&352&75&64 \\ \hline
0&6&1&0& 226&330&255&410&40&60 \\ \hline
\multicolumn{4}{||c||}{$\chi^2$(whole)/$\nu$}&
\multicolumn{1}{l}{7.97/5}& 
\multicolumn{1}{l||}{s.p.=0.17} & 
\multicolumn{1}{l}{8.34/5} & 
\multicolumn{1}{l||}{s.p.=0.15} & 
\multicolumn{1}{l}{13.92/5} & 
\multicolumn{1}{l||}{s.p=0.017} \\ 
\multicolumn{4}{||c||}{$\chi^2$(upward)/$\nu$}&
\multicolumn{1}{l}{7.52/3}& 
\multicolumn{1}{l||}{s.p.=0.068} & 
\multicolumn{1}{l}{4.90/3} & 
\multicolumn{1}{l||}{s.p.=0.18} & 
\multicolumn{1}{r}{5.78/3} & 
\multicolumn{1}{l||}{s.p=0.13} \\ 
\multicolumn{4}{||c||}{$\chi^2$(downward)/$\nu$}&
\multicolumn{1}{l}{4.29/3}& 
\multicolumn{1}{l||}{s.p.=0.24} & 
\multicolumn{1}{l}{3.84/3} & 
\multicolumn{1}{l||}{s.p.=0.28} & 
\multicolumn{1}{r}{8.18/3} & 
\multicolumn{1}{l||}{s.p=0.044} \\ \hline \hline
\multicolumn{4}{||c||}{zenith angle}&
\multicolumn{2}{|c||}{multi-GeV FC}&
\multicolumn{2}{|c||}{multi-GeV PC}&
\multicolumn{2}{|c||}{multi-GeV (FC+PC)} \\ \hline
$-1$&0&$-0$&6& 32&44&32&47&64&91 \\ \hline
$-0$&6&$-0$&2& 31&44&44&68&75&112 \\ \hline
$-0$&2&0&2& 45&64&91&131&136&195 \\ \hline
0&2&0&6& 73&78&68&124&151&202 \\ \hline
0&6&1&0& 49&98&66&83&115&181 \\ \hline
\multicolumn{4}{||c||}{$\chi^2$(whole)/$\nu$}&
\multicolumn{1}{l}{7.11/5}& 
\multicolumn{1}{l||}{s.p.=0.21} & 
\multicolumn{1}{l}{3.93/5} & 
\multicolumn{1}{l||}{s.p.=0.56} & 
\multicolumn{1}{l}{1.36/5} & 
\multicolumn{1}{l||}{s.p=0.93} \\ 
\multicolumn{4}{||c||}{$\chi^2$(upward)/$\nu$}&
\multicolumn{1}{l}{0.015/3}& 
\multicolumn{1}{l||}{s.p.=0.99} & 
\multicolumn{1}{l}{0.33/3} & 
\multicolumn{1}{l||}{s.p.=0.95} & 
\multicolumn{1}{l}{0.22/3} & 
\multicolumn{1}{l||}{s.p=0.97} \\ 
\multicolumn{4}{||c||}{$\chi^2$(downward)/$\nu$}&
\multicolumn{1}{l}{7.10/3}& 
\multicolumn{1}{l||}{s.p.=0.073} & 
\multicolumn{1}{l}{3.64/3} & 
\multicolumn{1}{l||}{s.p.=0.30} & 
\multicolumn{1}{l}{1.18/3} & 
\multicolumn{1}{l||}{s.p=0.76} \\ \hline \hline
\end{tabular}
\normalsize
\end{center}
\end{table}

It is generally known that 
a significance probability (s.p.) of $\ge$ 0.05 
in a statistical test indicates consistency for the two data sets. 
In the present cases, also, 
s.p. values of little larger than 0.05 
may indicate some degrees of mis-identification. 
In the following we carefully check 
the variances of the s.p. values of the respective distributions. 
In table \ref{table:chitab}, 
a small s.p. value of 0.068 was seen for 
the upward directed zenith angle distribution of sub-GeV $e$-like events. 
Further, in multi-GeV $e$-like events, 
we found an extremely rare s.p. value = 0.017 
for the whole zenith angle distribution. 
The downward directed events were also described by 
a s.p.= 0.044 which was smaller than 0.05, 
and the upward directed one of s.p.= 0.13 
was smaller than the other upward directed ones except 
for the  sub-GeV $e$-like events. 
Although multi-GeV $\mu$-like (FC + PC) events showed 
very large s.p. values for all the distributions, 
the downward-directed FC events could only be characterized by 
a s.p. value of 0.073. 
However, all distributions of the PC events show large s.p. values. 
These varying results might be connected to 
the identification problem of  muons and electrons.

For testing our s.p. values, 
we used the upward through-going muon flux and 
the upward-going stopping events in 
$-1 \le \cos \theta \le 0$ expressed with 10 bins 
for the former and 5 bins for the latter of the 79.3 kty exposure. 
Both these events are sure to be muons as they are irrelevant 
to the identification of muons and electrons. 
Similar $\chi^2$ tests were done for both distributions expressed 
by 5 bins each from the subdivision of the 79.3 kty exposure into 
the respective 33.0 kty and 46.3 kty exposures. 
From these data, we obtained s.p. values of 0.66 for 5 bins for 
the upward through-going muon flux, 
and a s.p. value of 0.37 for 5 bins in the upward-going stopping events. 
These s.p. values are not in conflict with being muons. 
From this we infer that the poor s.p. values 
shown in table \ref{table:chitab} 5, 
for some of the zenith angle distributions of 
$e$-like and of $\mu$-like events in the sub-GeV and multi-GeV ranges 
and also for the downward directed $\mu$-like (FC) events, 
may be connected to the poor identification capability for 
muons and electrons in the 33 kty and 46.3 kty exposures. 
Thus, we come to the conclusion that 
these events may show an appreciate mis-identification 
between muons and electrons. 
     
Accordingly, 
the comparison of the observed zenith angle distribution of 
various events with the expected one based on 
the neutrino oscillation hypothesis is still an open question. 
In addition, the comparisons of the upward-going muon flux with 
the null and the $\nu_{\mu} - \nu_{\tau}$ neutrino oscillation hypothesis 
have to face the problem about 
the absolute value of the neutrino and muon fluxes 
at ground level for different zenith angles. 
Namely, very recently, 
the measurement of atmospheric vertical muon fluxes 
at ground level in the momentum ranges 
from 200 MeV/$c$ to 120 GeV/$c$ by the CAPRICE group~\cite{JKre} 
and from 0.6 to 20 GeV/$c$ by the BESS group~\cite{MMo} 
were done by using low mass superconducting magnet spectrometers. 
The former are about 10 \% to 15 \% lower and the latter about 20 \% lower 
than the previous experimental results. 
The OKAYAMA group~\cite{STsu} using a solid iron magnet telescope 
has obtained the vertical muon fluxes from 1.5 to 100 GeV/$c$ 
which are consistent with both the above measurements. 
Also, the NMSU-WIZARD/CAPRICE98~\cite{PH} 
balloon-borne magnetic spectrometer reports new measurements 
at several atmospheric depths in the momentum range 0.3 - 20 GeV/$c$. 
 
\section{Conclusion}

Based on our MC simulation which includes fluctuations of 
the produced number of photoelectrons due to 
the stochastic nature of energy-loss mechanisms, 
we conclude that the total mis-identification rate for 
muon and electron events is larger than or equal to 20 \% 
in the sub-GeV region and also at least several \% in the multi-GeV region. 
In both cases the correct treatment of 
the particle identification would lead to 
an increase of the muon neutrino flux, 
namely of 20 \% in the sub-GeV range and 5 \% in the multi-GeV region. 
Among others, 
this mis-identification problem is related to the fact 
that electrons will always initiate electromagnetic cascades 
leading to a fuzzy image of the \v{C}erenkov ring pattern. 
On the other hand muons can occasionally undergo 
electromagnetic interactions 
(knock-on production, bremsstrahlung and direct electron pair production). 
In these cases the  muon track overlaps with the shower initiated by 
the secondaries created by the muon. 
The emerging pattern can easily be  mistaken for an electron.

In addition, our $\chi^2$ tests showed that 
the zenith angle distributions observed by Super K. 
for $e$-like and $\mu$-like (FC), except $\mu$-like (PC) events 
in the first half of the total exposure of 79.3 kty (33.0 kty) 
and the second half (46.3 kty) are inconsistent both 
for the sub-GeV and multi-GeV ranges.
In special, very different s.p. values between 0.99 and 0.073 
for both observed zenith angle distributions of 
upward-directed and downward-directed $\mu$-like (FC) events suggest 
that the downward going events suffer from mis-identifications 
between muons and electrons. 
Given these results, 
we conclude that due to the uncertainty of the identification capability 
for muon and electron events in the Super K. detector 
one cannot yet firmly establish the existence of 
an atmospheric neutrino anomaly. 
A better knowledge and precision treatment of 
the particle identification technique including discussions of 
the absolute intensities for the zenith angle dependent 
upward going events and the accurate atmospheric neutrino fluxes 
in general will certainly help to resolve the remaining problems. 

\ack

We wish to express our thanks to Professor Claus Grupen of 
University of Siegen for polishing the English sentences.

 \end{document}